\documentstyle[12pt,amsmath,epsfig,psfrag]{article}

\newcommand{\f}{\frac}
\newcommand{\non}{\nonumber \\}
\newcommand {\beq}{\begin{equation}}
\newcommand {\eeq}{\end{equation}}
\newcommand {\beqa}{\begin{eqnarray}}
\newcommand {\beqal}{\begin{eqnarray}\label}
\newcommand {\eeqa}{\end{eqnarray}}
\newcommand {\bc}{\begin{center}}
\newcommand {\ec}{\end{center}}
\newcommand {\s}{\sigma}
\newcommand {\al}{\alpha^{'}}
\newcommand {\als}{\alpha^{'2}}
\newcommand {\alp}{\alpha}

\newcommand {\vth}{\vartheta}
\newcommand {\ka}{\kappa}
\newcommand {\pa}{\partial}

\newcommand {\De}{\Delta}
\newcommand {\ep}{\epsilon}
\newcommand {\kpa}{k_{\parallel}}
\newcommand {\kpai}{k_{\parallel i}}
\newcommand {\kpaj}{k_{\parallel j}}
\newcommand {\kpe}{k_{\perp}}
\newcommand {\Tr}{\mbox{Tr}}

\newcommand{\ket}[1]{\left | #1 \right\rangle}
\newcommand{\expt}[1]{\left\langle #1 \right\rangle}

\newcommand {\w}{\wedge}
\newcommand {\noi}{\noindent}

\newcommand{\alpb}
{\left[\begin{smallmatrix}
\alp \\ \beta 
\end{smallmatrix}\right]}
\def\nn{\nonumber}
\newcommand{\oo}
{\left[\begin{smallmatrix}
1/2 \\ 1/2
\end{smallmatrix}\right]}
\def\nn{\nonumber}

\def\nn{\nonumber}

\begin{document}

\title{
\hfill\parbox{4cm}{\normalsize IP/BBSR/2006-08}\\
\vspace{1cm}
Closed string exchanges on $C^2/Z_2$ in a background $B$-field
\author{Swarnendu Sarkar \footnote{email:swarnen@iopb.res.in}\\
\\
{\em Institute of
Physics, Bhubaneswar}\\
{\em India 751 005}}}
\maketitle

\begin{abstract}
In an earlier work it was shown that the IR singularities arising in the nonplanar one loop two point function of a noncommutative ${\cal N}=2$ gauge theory can be reproduced exactly from the massless closed string exchanges. The noncommutative gauge theory is realised on a fractional $D_3$ brane localised at the fixed point of the $C^2/Z_2$ orbifold. In this paper we identify the contributions from each of the closed string modes. The sum of these adds upto the nonplanar two-point function.  
\noindent
\end{abstract}
\newpage

\section{Introduction}

A generic feature of noncommutative field theories is the coupling of the ultraviolet and the infrared sectors. These theories can be realised as low energy limits of open strings in constant background $B$-field, in the Seiberg-Witten limit \cite{Seiberg}. The loop integrals in the nonplanar diagrams of the noncommutative theories are regulated in the ultraviolet but are divergent when the external momentum goes to zero. This has a natural interpretation in terms of open closed string channel duality where the UV region of the open string channel can be mapped to the IR of the closed string. 

As a consequence of the background $B$-field, the open and closed strings couple to different metrics on the brane and the bulk respectively.
We recall that in the presence of a constant background $B$-field that is nonzero only for the directions $(i,j)$ that are along the brane, the open string modes couple to the metric $G$. This is related to the metric $g$ that couples to the closed string modes by \footnote{We will use capital letters $(M, N,...)$ to denote
general spacetime indices and small letters $(i,j, ...)$ for coordinates
along the $D$-brane.} 

\beqal{gth}
G^{MN}&=&\left(\f{1}{g+2\pi\al B}g\f{1}{g-2\pi\al B}\right)^{MN}\non
G_{MN}&=&g_{MN}-(2\pi\al)^2(Bg^{-1}B)_{MN}
\eeqa

Noncommutative field theory arises in the following Seiberg-Witten limit, 

\beqal{swl}
\al \sim \epsilon^{1/2} \rightarrow 0
\mbox{\hspace{0.1in};\hspace{0.1in}}
g_{ij} \sim \epsilon \rightarrow 0
\eeqa

A framework in which the phenomenon discussed in the opening paragraph can be 
analysed is the the world-sheet open closed string duality in the presence of 
a background $B$-field. It is well known that an open string one loop 
amplitude can be seen as tree-level exchanges of closed string modes. 
The circulating modes in the loop degenerates to the massless ones in the 
open string channel ($t \rightarrow \infty$) and to the closed string 
massless modes for ($t \rightarrow 0$). Where $t$ is the modulus of the one 
loop cylinder diagram . Again the UV region of the open string modes 
correspond to the IR due to the massless closed string states. The 
correspondence here is between the full tower of open string modes and the 
massless closed string modes. It would be interesting to analyse situations 
in which the correspondence is exact between massless modes on either side. 
The problem outlined earlier in the context of noncommutative gauge theory is 
analogous to this situation. We have studied this along these lines in the 
context of 
the bosonic string theory in \cite{myopcl1}. 
Though we do not expect an exact correspondence in this bosonic model, 
the analysis has led to various fruitful insights. An exact matching was 
shown for the ${\cal N}=2$ gauge theory that is realised on a fractional 
$D_3$ brane localised at the fixed point of $C^2/Z_2$ orbifold and the bulk 
closed string theory \cite{myopcl2}. The hypermultiplets in this gauge theory 
are projected out and the theory is non-conformal. It was further noticed 
that the role played by the $B$-field is essentially that of a regulator. 
Thus leading us to conclude that the UV/IR singularities in noncommutative 
gauge theory can be seen as IR of the massless closed string exchanges 
whenever the correspondence between the massless states hold for the 
commutative models perturbatively. The fact that this is true for the 
ordinary ${\cal N}=2$ theory was observed in \cite{douglas2}.  
A class of such models were later analysed in 
\cite{lerda1,Polchinski:2000mx,lerda2,DiVecchia}. Also see 
\cite{Bertolini:2003iv} for reviews.

In this paper, we study the massless closed string exchanges on $C^2/Z_2$ 
in a background $B$-field. We identify the contribution from each of the 
massless modes. The procedure followed is same as that in \cite{myopcl1}. 
We first derive the couplings for the gauge field with the massless closed 
string modes from the DBI and the Chern-Simons action. We then compute the 
nonplanar two point function with two gauge field insertions which then adds 
upto the one-loop amplitude computed from string theory.

This paper is organised as follows. In Section 2 we review the nonplanar 
one-loop open string two point amplitude in the closed string channel. 
In Section 3 we study the closed string exchanges. We first study the massless 
closed string exchanges for flat space background in Section 3.1. This 
amplitude vanishes as we can also see from the one loop string computation. 
Massless exchanges on  $C^2/Z_2$ is the studied in Section 3.2. 
We end with conclusions in Section 4.

\section{Review of one loop amplitude}

In this section we review the results of the one loop open string 
amplitude on the $C^2/Z_2$ orbifold with two gauge field insertions. Here 
we give a brief outline of the computation, the 
details of which may be found in \cite{myopcl2}. 

\beqal{2pt}
A(p,-p)&=&iV_4
\det(g+2\pi\al B)\int^{\infty}_{0}\f{dt}{4t}(8\pi^2\al t)^{-2}
\times \non &\times&\sum_{(\alp,\beta,g_i)}Z {\alpb}_{g_i}
\int_{0}^{2\pi t}dy\int_{0}^{2\pi t}dy^{'}
\expt{V(p,x,y)V(-p,x^{'},y^{'})}_{(\alp,\beta)}
\eeqa

\noindent
The factor of $\det(g+2\pi\al B)$ comes from the trace over the 
world sheet bosonic zero modes.
The sum over $(\alp,\beta)=(0,1/2)$ corresponds to spin structures $(\alp,\beta)=(0,1/2)$ corresponding 
to the NS-R sectors 
and the GSO projection, whereas the sum over ${g_i}$ projects onto states invariant under the the orbifold action. The elements 
$Z_{g_i} {\alpb}$ are the traces over the states computed in \cite{myopcl2}. 
The vertex operator $V(p,x,y)$ is given by,

\beqa
V(p,x,y)=\f{g_o}{(2\al)^{1/2}}\ep_j\left(i\pa_y X^j+4p.\Psi\Psi^j\right)
e^{ip.X}(x,y)
\eeqa

\noindent
For the flat space, it is well known that amplitudes with less that four 
boson insertions vanish. However, in this model the two point amplitude 
is nonzero. We will now compute this amplitude in the presence of 
background $B$-field. First note that the bosonic correlation function, 
$\expt{:\pa_yX^ie^{ip.X}::\pa_y^{'}X^ie^{-ip.X}:}$, does not 
contribute to 
the two point amplitude as it is independent of the spin structure. The 
two point function would involve the sum over the $Z_{g_i} \alpb$ which 
makes this contribution zero as the vacuum amplitude vanishes. The nonzero 
part of the amplitude will be obtained 
from the fermionic part,

\beqa
\ep_k\ep_l\expt{:p.\Psi\Psi^ke^{ip.X}::
p.\Psi\Psi^le^{-ip.X}:}&=&\ep_k\ep_l p_ip_j
\left(G^{il}G^{jk}-G^{ij}G^{kl}\right)\times\\
&\times&{\cal G}^2\alpb (w-w^{'}) \expt{:e^{ip.X}::e^{-ip.X}:}\nn
\eeqa

\noindent
For the planar two point amplitude, both the vertex operators would be 
inserted at the same end of the cylinder (i.e. at $w=0+iy$ or $\pi+iy$). 
In this case, the sum in the two point amplitude reduces to,

\beqal{sum}
\sum_{(\alp,\beta,g_i)}Z\alpb_{g_i}{\cal G}^2\alpb(i\De y/2\pi)&=&
\sum_{(\alp,\beta)}Z\alpb_e{\cal G}^2\alpb(i\De y/2\pi)\non &+&
\sum_{(\alp,\beta)}Z\alpb_g{\cal G}^2\alpb(i\De y/2\pi)\nn
\eeqa
\beqa
&=&\f{4\pi^2}{\eta(it)^{6}\vth^2_1(i\De y/2\pi,it)}\sum_{(\alp,\beta)}
\vth^2(0,it)\alpb\vth^2\alpb(i\De y/2\pi,it)+ \\
&+& \f{16\pi^2}{\vth^{2}_{1}(i\De y/2\pi,it)
\vth^2_2(0,it)}\left[\vth^2_3(i\De
y/2\pi,it)\vth^{2}_{4}(0,it)-
\vth^2_4(i\De y/2\pi,it)\vth^{2}_{3}(0,it)\right]\nn
\eeqa

\noindent
where, $\De y=y-y^{'}$.
We have separated the total sum as the sum over the two $Z_2$ 
group 
actions. In writing this out we have used the following identity 

\beqa\eta(it)=\left[\f{\pa_{\nu}\vth_1(\nu,it)}{-2\pi}\right]^{1/3}_{\nu=0}
\eeqa

\noindent
Now, the first term vanishes due to the following identity

\beqa
\lefteqn{\sum_{(\alp,\beta)}\vth\alpb (u)\vth\alpb (v)\vth\alpb (w)\vth\alpb (s)=}\\ & & 2\vth\oo (u_1)\vth\oo (v_1)\vth\oo (w_1)\vth\oo (s_1)\nn
\eeqa

\noindent
where,
\beqa
u_1&=&\f{1}{2}(u+v+w+s) \mbox{\hspace{0.2in}} v_1=\f{1}{2}(u+v-w-s) 
\mbox{\hspace{0.2in}}\non w_1&=&\f{1}{2}(u-v+w-s) \mbox{\hspace{0.2in}}
s_1=\f{1}{2}(u-v-w+s)
\eeqa

\noindent
and noting that, $\vth\oo(0,it)=0$,
in the same way as the flat case that makes amplitudes with two 
vertex insertions vanish. The second term is a constant also due to,

\beqal{ident}
\vth^2_4(z,it)\vth^2_3(0,it)-\vth^2_3(z,it)\vth^2_4(0,it)
=\vth^2_1(z,it)\vth^2_2(0,it)
\eeqa

\noindent
For the nonplanar amplitude, which we are ultimately interested in, we 
need to put the two vertices at the two ends of the cylinder such that, 
$w=\pi+iy$ and $w^{'}=iy^{'}$. It can be seen that the fermionic part
of the correlator is constant and independent of $t$, same as the 
planar case following from the identity (\ref{ident}). The effect of 
nonplanarity 
and the regulation of the two point function due to the background 
$B$-field is encoded in the correlation functions for the exponentials. 
To focus on the closed string exchanges, we study the contribution $t \rightarrow 0$ limit of this amplitude. Finally the two point function reduces to,

\beqal{2ptfinal}
A(p,-p)=iV_4\det(g+2\pi\al B)\left(\f{g_o^2}{8\pi^2\al}\right)
\ep_k\ep_l p_ip_j
\left(G^{il}G^{jk}-G^{ij}G^{kl}\right)I(p)
\eeqa

where,
\beqal{ip}
I(p)&=&\int ds s^{-1}
\exp\left\{-\f{\al\pi s}{2}p_ig^{ij}p_j\right\}\non
&=&4\pi\int \f{d^2\kpe}{(2\pi)^2}\f{1}{\kpe^2+p_ig^{ij}p_j}
\eeqa

\noindent
Note that the integral is written in terms of $s=1/t$ 
that is again
rewritten as an integral over $\kpe$, the momentum in the
directions transverse to the brane for
closed strings. The nonzero contribution to the two point amplitude
in (\ref{2pt}) comes from the $\Tr_{NS}\left[gq^{L_0}\right]$ and
$\Tr_{NS}\left[g(-1)^F q^{L_0}\right]$. These correspond to 
anti-periodic NS-NS and periodic R-R
closed strings in the twisted sectors respectively. The fractional
$D_3$-brane is localised at the fixed point of $C^2/Z_2$. Thus the
twisted sector closed string states that couple to it 
are localised at the fixed point and are free to move in the
six directions transverse to the orbifold. This is the origin of the
momentum integral (\ref{ip}) in two directions transverse to the
$D$-brane. Before going into the calculation for the closed string 
exchanges let us break here to discuss the spectrum for the closed 
strings on this orbifold.

The closed string theory consists of additional twisted sectors apart 
from the untwisted sectors. The orbifold action an the space-time 
implies the following boundary conditions on the world-sheet bosons and 
fermions,

\beqa
X^{I}(\s+2\pi,\tau)&=&\pm X^{I}(\s,\tau)\non
\psi^{I}(\s+2\pi,\tau)&=&\pm \psi^{I}(\s,\tau)
\mbox{\hspace{.2in}} I=6,7,8,9
\eeqa

For the world sheet fermions, the $(+)$-sign stands for the NS-sector 
and the $(-)$-sign for the R-sector. For the other directions the boundary 
conditions on the world-sheet fields are as usual. We will first list 
the fields in the untwisted sector. In the NS-NS sector the massless 
states invariant under the orbifold projection are,

\beqa
\psi_{-1/2}^{I}\tilde{\psi}_{-1/2}^{J}\ket{0,k}  
\eeqa

where, $I,J=\{2,3,4,5\}$ or $I,J=\{6,7,8,9$\}. The first set of 
oscillators give the graviton, antisymmetric 2-form field, and the 
dilaton. The second set gives sixteen scalars.

The orbifold action on the spinor of $SO(8)$ is given by,

\beqa
\ket{s_1,s_2,s_3,s_4} \rightarrow 
e^{i\pi(s_3+s_4)}\ket{s_1,s_2,s_3,s_4}
\eeqa

The $Z_2$ invariant R-R state is formed by taking both the left 
the right states to be either even or odd under $Z_2$ projection
corresponding to $s_3+s_4=0$ or $s_3+s_4=\pm 1$ respectively. GSO 
projection, restricting to both the left and right states to be of the 
same chirality gives thirty two states. These states correspond to four 
2-form fields and eight scalars.

Let us now turn to the twisted sectors. 
For the twisted sectors the ground state energy for both the
NS and the R sectors vanish. In the NS sector the massless modes
come from $\psi^I_0$, $I=6,7,8,9$ oscillators which form a spinor
representation of $SO(4)$. With the GSO and the orbifold projections,
the closed string spectrum is given by, $2\times 2 =[0]+[2]$. The [0]
and the self-dual [2] constitute the four massless scalars in the
NS-NS
sector. Similarly, in the R sector, the massless modes are given by
$\psi^I_0$ for
$I=2,3,4,5$. Thus giving a scalar and a two-form self-dual field in the
closed string R-R sector. The twisted states can also be seen as arising from
the dimensional reduction of $p$-form fields on a vanishing 2-cycle. We will see this more elaborately in Section (\ref{orbifold}). We will write down the couplings of these twisted and untwisted fields from the DBI and the Chern-Simons action and then compute the contribution from each of these modes, the sum of which would reproduce (\ref{2ptfinal}).

\section{Closed string exchanges}

After reviewing the analysis of the nonplanar two point amplitude, let us now proceed to study in detail the massless closed string exchanges. 
In this section we will calculate the contribution to the nonplanar two point function with two gauge fields on the brane with massless closed string exchanges coming from the NS-NS and the R-R sectors. As in \cite{myopcl1}, we will calculate these in three different limits of the closed string metric.

\begin{enumerate}

\item In this case the background $B$-field 
is assumed to be small and the closed string metric, $g=\eta$.
The amplitude will be analysed to ${\cal O}(B^2)$.

\item The Seiberg Witten limit when $g=\ep\eta$ with the amplitude expanded to
 ${\cal O}(\ep^2/(2\pi\al)^2)$.

\item The case when the open
string metric on the brane, $G=\eta$ so that $g=-(2\pi\al)^2B^2 +{\cal 
O}(\alpha^{'4})$ and the amplitude will be expanded to
${\cal O}((2\pi\al)^2)$

\end{enumerate}

\subsection{Type IIB on flat space}\label{flat}

We will start by considering the flat 10D case and consider massless closed string exchanges for a $D_3$ brane. The two point function will be shown to vanish as expected from the analysis in the previous section. See eqn(\ref{sum}). The results here will be necessary for the later part of this section when we study the exchanges on $C^2/Z_2$ orbifold. These will precisely be the contributions from the untwisted states upto an overall constant.
To begin we write down the supergravity action for the type IIB theory
in the Einstein frame \footnote{$(A_p\w B_q)_{i_1...i_p j_1...j_q}=\f{(p+q)!}{p!q!}A_{[i_1...i_p}B_{j_1...j_q]}$. $A_p=\f{1}{p!}\omega_{i_1...i_p}dx^{i_1}\w...\w dx^{i_p}$ and $*A_p=\f{1}{(d-p)!}\omega_{i_1...d_p}\ep_{j_1..j_{d-p}}^{i_1...d_p}dx^{j_1}\w...\w dx^{j_{d-p}}$} 

\beqal{f2b}
S_{IIB}&=&\f{1}{2\ka^2}\left[\int d^{10}x 
\sqrt{-g}R-\f{1}{2}\int\left[d\phi\w*d\phi
+e^{-\phi}H_3\w*H_3\right]\right]\non
&-&\f{1}{4\ka_{10}^2}\left[\int e^{2\phi}F_1\w*F_1+e^{\phi} 
F_3\w*F_3+\f{1}{2}F_5\w*F_5\right]+...
\eeqa

\noi
where $\ka^2=\ka^2_{10}e^{-2\phi_0}$, and

\beqa
H_3=db \mbox{\hspace{.2in}} F_1=dC_0 \mbox{\hspace{.2in}} F_3=dC_2 \mbox{\hspace{.2in}} F_5=dC_4
\eeqa

\noi
Where $b$ is the two form antisymmetric NS-NS field \footnote{We will denote 
the constant part of the NS-NS two form field as $B$ and the fluctuation about this as $b$. The same field on the brane will be identified as the field strength of the $U(1)$ gauge field}. We have omitted the other terms in the action (\ref{f2b}) as we are only interested in the propagators for the closed string modes that will 
be needed to compute the two point amplitude in the later part of this 
section. We first write down the propagators for the NS-NS modes that have 
been worked out in \cite{myopcl1} in the context of bosonic string theory.

For the dilaton we have,

\beqal{dp}
\expt{\phi\phi}=-2i\ka^2\f{1}{\kpe^2+g^{ij}\kpai\kpaj} 
\eeqa

and for the propagating $b$ field,

\beqal{bp}
\expt{b_{IJ}b_{I^{'}J{'}}}=-\f{2i\ka^2}{(2\pi\al)^2}\f{g_{I[J{'}}g_{I^{'}]J}}
{\kpe^2+g^{ij}\kpai\kpaj}
\eeqa
 
Note that the factor of $1/(2\pi\al)^2$ in the $b$-field propagator has 
been included as the sigma model in the noncommutative description is 
defined with $(2\pi\al)B$ coupling. Similarly for the graviton,

\beqal{gp}
\expt{h_{IJ}h_{I^{'}J{'}}}=-2i\ka^2\f{[\eta_{I\{J{'}}\eta_{I^{'}\}J}-2/(D-2)
\eta_{IJ}\eta_{I^{'}J{'}}]}{\kpe^2+\kpa^2}
\eeqa

The R-R modes that will be relevant for our discussions are the zero-form, $C_0$ and the two-form, $C_2$.
For the R-R modes, the propagators are same as that of the NS-NS modes upto normalisations,

\beqal{rrp}
\expt{C_0C_0}&=&\f{\ka_{10}^2}{\ka^2}\expt{\phi\phi}\non
\expt{C_{2IJ}C_{2I^{'}J^{'}}}&=&\f{\ka_{10}^2}{\ka^2}
(2\pi\al)^2\expt{b_{IJ}b_{I^{'}J{'}}}
\eeqa

In the following analysis we shall restrict the propagators to the 
values for the closed string metric $g^{ij}$ in the various limits stated at the begining of this section.  We are interested in the correction to the quadratic term in the
effective action for the
gauge field on the brane. This can be constructed from the vertices
and the propagators for the intermediate massless closed
string states. The correction for the nonplanar diagram can be written
as,

\beqal{pos2pt}
A_2(bb)=\int d^{p+1}\xi\int d^{p+1}\xi^{'}b(\xi)b(\xi^{'})
V<\chi(\xi)\chi(\xi^{'})>V
\eeqa
\noindent
where,
\beqa
<\chi(\xi)\chi(\xi^{'})>=\int
\f{d^Dk}{(2\pi)^D}<\chi(\kpe,\kpa)\chi(-\kpe,-\kpa)>e^{-i\kpa(\xi-\xi^{'})}
\eeqa

\noindent
Where $\kpe$ is the component of momentum of the closed string mode perpendicular to the brane and $\kpa$ parallel to it. 
We can rewrite eqn(\ref{pos2pt}) in momentum space coordinates as,

\beqal{eff}
A_2(bb)&=&V_{p+1}\int \f{d^{p+1}p}{(2\pi)^{p+1}}b(p)b(-p)\int
\f{d^l\kpe}{(2\pi)^l}V<\chi(\kpe,-p)\chi(-\kpe,p)>V\non
&=&V_{p+1}\int \f{d^{p+1}p}{(2\pi)^{p+1}}b(p)b(-p)L_2(p,-p)
\eeqa

\noindent
Where $l=D-(p+1)$ is the number of directions transverse to the brane.
In the planar two point function, both the vertices are on the same end
of the
cylinder in the world-sheet computation. In the field theory this
corresponds to putting both the
vertices at the same position on the $D$-brane. In other words,
in the expansions of the DBI and Chern-Simons action, we should be looking 
for $b^2\chi$
vertices on one end and a $\chi$ tadpole on the
other. In this case, from the above calculation, $\kpa=0$. So
the closed string propagator is just $1/\kpe^2$, i.e. the
propagator is not modified by the momentum of the gauge field on the
brane. This is what we expect, as in the field theory on the brane, the
loop integrals are not modified for the planar diagrams. Here we will
only concentrate on the nonplanar sector. Finally to compare with the string theory amplitude, we will identify,

\beqal{identify}
b_{kl}(p)\equiv \f{g_0}{\sqrt{2\al}}F_{kl}(p)
=\f{g_0}{\sqrt{2\al}}p_{[k} A_{l]}(p)
\eeqa

\subsubsection{NS-NS exchange}\label{fns}

We now turn to the DBI and 
the Chern-Simons action of a $D_3$ brane for calculating the 
massless closed string couplings to the gauge field

\beqal{fdbi}
S_{DBI}=-\tau_3\int d^{4}\xi 
\sqrt{g+2\pi\al(B+b)e^{-\f{\phi}{2}}}
\eeqa

$g$ is the closed string metric in the Einstein frame, $B$ is
the constant two form
background field and $b$ is the fluctuation of the two form field. As 
mentioned earlier, the
$b$-field on the brane is interpreted as the two form field strength
for the $U(1)$ gauge field and in the bulk it is the usual two form
potential.
The NS-NS field content is same as that of the bosonic theory. To compute 
the amplitude due to the exchange of these fields we need to set 
$D\rightarrow 10$ in the amplitudes calculated in \cite{myopcl1}. We recollect 
these expressions for the three cases here. For small $B$ expansion we have,

\beqal{ffinal1}
L_2&=&-i\ka^2\tau_3^2(2\pi\al)^2\int\f{d^6k}{(2\pi)^6}
\f{1}{\kpe^2+p^2}\times\\&\times&[\f{(2\pi\al)^2}{4}B^{kl}B^{k^{'}l^{'}}
+\f{1}{4}\left[1-\f{(2\pi\al)^2}{2}\Tr(B^2)\right]\left(\eta^{ll^{'}}\eta^{kk^{'}}-
\eta^{lk^{'}}\eta^{kl^{'}}\right)\non
&+&\f{(2\pi\al)^2}{2}\left[(B^2)^{ll^{'}}\eta^{kk^{'}}-
(B^2)^{lk^{'}}\eta^{kl^{'}}\right]+(kl)\leftrightarrow(k^{'}l^{'})]\nonumber
\eeqa

\noi
The contribution to (\ref{ffinal1}) comes from the graviton, dilaton and the 
propagating antisymmetric $b$ field in the bulk.
For the noncommutative limit, $g=\ep\eta$,

\beqal{ffinal2}
L_2&=&-i\mbox{det}(2\pi\al B)\ka^2\tau_3^2
\int\f{d^6\kpe}{(2\pi)^6}\f{1}{\kpe^2+\ep^{-1}p^2}
[{\cal O}(1)+{\cal O}(\ep^2)]
\eeqa

\noi
where,

\beqa
{\cal O}(1)=
\left[\f{1}{4}
\left(\f{1}{B}\right)^{kl}\left(\f{1}{B}\right)^{k^{'}l^{'}}
+(kl) \leftrightarrow (k^{'}l^{'})\right]
\eeqa

\beqa
{\cal O}(\ep^2)&=&
\f{\ep^2}{(2\pi\al)^2}\f{1}{2}\left[\left[\left(\f{1}{B^3}\right)^{kl}
-\f{1}{4}\Tr\left(\f{1}{B^2}\right)\left(\f{1}{B}\right)^{kl}\right]
\left(\f{1}{B}\right)^{k^{'}l^{'}}\right]\non
&+&\f{\ep^2}{(2\pi\al)^2}\left[
\f{1}{4}\left(\f{1}{B^2}\right)^{kk^{'}}\left(\f{1}{B^2}\right)^{ll^{'}}
-\f{1}{4}\left(\f{1}{B^2}\right)^{k^{'}l}\left(\f{1}{B^2}\right)^{kl^{'}}
\right]\non
&+& (kl) \leftrightarrow (k^{'}l^{'})
\eeqa

\noi
For the noncommutative limit, $G=\eta$,

\beqal{ffinal3}
L_2=-i\det(2\pi\al
B)\ka^2\tau_3^2\int\f{d^6\kpe}{(2\pi)^6}
\f{1}{\kpe^2+\tilde{p}^2(2\pi\al)^2}[{\cal
O}(1)
+{\cal O}(\als)]
\eeqa

\beqa
{\cal O}(1)=
\left[\f{1}{4}
\left(\f{1}{B}\right)^{kl}\left(\f{1}{B}\right)^{k^{'}l^{'}}
+(kl) \leftrightarrow (k^{'}l^{'})\right]
\eeqa

\beqa
{\cal O}(\als)&=&
(2\pi\al)^2\f{1}{2}\left[\left[B^{kl}
-\f{1}{4}\Tr(B^2)\left(\f{1}{B}\right)^{kl}\right]
\left(\f{1}{B}\right)^{k^{'}l^{'}}\right] \non
&+&(2\pi\al)^2\left[
\f{1}{4}\left(\eta^{ll^{'}}\eta^{kk^{'}}
-\eta^{kl^{'}}\eta^{lk^{'}}\right)
\right]\non
&+& (kl) \leftrightarrow (k^{'}l^{'})
\eeqa

For the noncommutative limits we get contributions only from the dilaton 
and the antisymmetric $b$ field. The graviton does not contribute to the 
order to which we are working here.

\subsubsection{R-R exchange}\label{frr}
The R-R couplings will be given by the usual Chern-Simons terms. We will consider here the commutative description of these terms. For a discussion of noncommutative description see \cite{mukhi,liu}

\beqal{fcs}
S_{CS}=i\mu_3\int_4\sum_n C_n\w e^{2\pi\al(B+b)}
\eeqa

\noi
Expanding (\ref{fcs}) and picking out the forms proportional to the volume 
form with one $b$ insertion we get

\beqal{fcse}
S_{CS}=i\mu_3\left[(2\pi\al)^2\int_4 C_0 B\w b + (2\pi\al)\int_4 C_2\w b 
\right]
\eeqa
\\
\noi
\underline{{\bf\it $C_2$ Exchange}} :
\\
\\
\noi
The coupling of $C_2$ to $b$ is given by,

\beqal{vc21}
V_{bC_2}=\f{i\mu_3}{4}(2\pi\al)\ep^{ijkl}
\eeqa

The two point amplitude can be worked out as in (\ref{eff}). Using the 
propagator (\ref{rrp}), for small $B$ expansion and $g=\eta$, the contribution to the two point amplitude is,

\beqal{fc21}
L_2(bC_2b)&=&\f{i}{2}\ka_{10}^2\mu_3^2(2\pi\al)^2\int
\f{d^6\kpe}{(2\pi)^6}\f{1}{\kpe^2+p^2}
\left[\f{1}{2}\left(\eta^{kk^{'}}\eta^{ll^{'}}-\eta^{kl^{'}}\eta^{lk^{'}}\right)\right]\non
&+& (kl) \leftrightarrow (k^{'}l^{'})
\eeqa

For the noncommutative cases, we will rewrite the coupling (\ref{vc21}) as 
\beqal{vc22}
\lefteqn{V_{bC_2}=\sqrt{2\pi\al 
B}\f{i\mu_3}{32(2\pi\al)}\left(\f{1}{B}\right)^{pq}
\left(\f{1}{B}\right)^{rs} \ep_{pqrs}\ep^{ijkl}}\\
& &= \sqrt{2\pi\al B}\f{i\mu_3}{4(2\pi\al)}\left[\left(\f{1}{B}\right)^{ik}
\left(\f{1}{B}\right)^{jl}-\left(\f{1}{B}\right)^{jk}\left(\f{1}{B}\right)^{il}-
\left(\f{1}{B}\right)^{ij}\left(\f{1}{B}\right)^{kl}\right]\nn
\eeqa
\\
\\
Where we have used the fact that, for an 
antisymmetric matrix $M$ of rank $2n$, 
$\sqrt{M}=\f{(-1)^n}{2^n n!}\ep_{\mu_1..\mu_{2n}}M^{\mu_1\mu_2}...M^{\mu_{2n-1}
\mu_{2n}}$. We can now calculate the two point function. 
Note that the dependence of the amplitude on the closed string metric $g$ only comes from the propagator. For $g=\ep\eta$ 
this is given by,

\beqal{fc22}
L_2(bC_2b)=
&=&i\mbox{det}(2\pi\al
B)\ka_{10}^2\mu_3^2\f{\ep^2}{(2\pi\al)^2}\int
\f{d^6\kpe}{(2\pi)^6}\f{1}{\kpe^2+\ep^{-1}p^2}
\times \non
&\times&[\left[\f{1}{2}\left(\f{1}{B^3}\right)^{kl}
-\f{1}{8}\Tr\left(\f{1}{B^2}\right)\left(\f{1}{B}\right)^{kl}\right]
\left(\f{1}{B}\right)^{k^{'}l^{'}} \non
&+&\f{1}{4}\left[\left(\f{1}{B^2}\right)^{kk^{'}}\left(\f{1}{B^2}\right)^{ll^{'}}
-\left(\f{1}{B^2}\right)^{k^{'}l}\left(\f{1}{B^2}\right)^{kl^{'}}\right]
\non
&+& (kl) \leftrightarrow (k^{'}l^{'})]
\eeqa

and for $G=\eta$

\beqal{fc23}
L_2(bC_2b)&=&
i\mbox{det}(2\pi\al B)\ka_{10}^2\mu_3^2(2\pi\al)^2
\int \f{d^6\kpe}{(2\pi)^6}
\f{1}{\kpe^2+\tilde{p}^2/(2\pi\al)^2}\times\non &\times&
\left[\left[\f{1}{2}B^{kl}-\f{1}{8}\Tr(B^2)\left(\f{1}{B}\right)^{kl}\right]
\left(\f{1}{B}\right)^{k^{'}l^{'}}
+\f{1}{4}\left(\eta^{ll^{'}}\eta^{kk^{'}}
-\eta^{kl^{'}}\eta^{lk^{'}}\right)\right]\non &+& (kl) \leftrightarrow
(k^{'}l^{'})
\eeqa
\\
\\

\noi
\underline{{\it $C_0$ Exchange}} :\\
\\
\noi
Reading from (\ref{fcse}) the coupling of $C_0$ to the gauge field on the brane is given by

\beqa
V_{bC_0}=\f{i\mu_3}{4}(2\pi\al)^2 B_{ij}\ep^{ijkl}
\eeqa

\noi
The noncommutative couplings can be obtained as the $C_2$ case, and 
finally contracting the answer with $B_{ij}$. This gives

\beqa
V_{bC_0}=\sqrt{2\pi\al B}\f{i\mu_3}{2}\left(\f{1}{B}\right)^{kl}
\eeqa

And for the two point function, we have

\beqal{fc01}
L_2(bC_0b)&=&i\ka_{10}^2\mu_3^2 (2\pi\al)^4 \int\f{d^6\kpe}{(2\pi)^6}
\f{1}{\kpe^2+p^2}\times\\&\times&
[\f{1}{4}B^{kl}B^{k^{'}l^{'}}-\f{1}{8}\Tr(B^2)\left[\eta^{kk^{'}}\eta^{ll^{'}}-\eta^{kl^{'}}\eta^{lk^{'}}\right]\non
&+&\f{1}{2}\left[\eta^{kk^{'}}(B^2)^{ll^{'}}-\eta^{kl^{'}}(B^2)^{lk^{'}}\right]
+(kl) \leftrightarrow (k^{'}l^{'})]
\mbox{\hspace{.1in} (for small $B$)}\nn
\eeqa

\beqal{fc02}
L_2(bC_0b)&=&i\mbox{det}(2\pi\al)\ka_{10}^2\mu_3^2\int \f{d^6\kpe}{(2\pi)^6}
\f{1}{\kpe^2+\ep^{-1}p^2}\times\non&\times&\left[\f{1}{4}
\left(\f{1}{B}\right)^{kl}\left(\f{1}{B}\right)^{k^{'}l^{'}}
+(kl) \leftrightarrow (k^{'}l^{'})\right]\mbox{\hspace{.1in} (for $g=\ep\eta$)}
\eeqa

\beqal{fc03}
L_2(bC_0b)&=&i\mbox{det}(2\pi\al)\ka_{10}^2\mu_3^2\int \f{d^6\kpe}{(2\pi)^6}
\f{1}{\kpe^2+\tilde{p}^2/(2\pi\al)^2}\times\non&\times&\left[\f{1}{4}
\left(\f{1}{B}\right)^{kl}\left(\f{1}{B}\right)^{k^{'}l^{'}}
+(kl) \leftrightarrow (k^{'}l^{'})\right] \mbox{\hspace{.1in} (for $G=\eta$)}
\eeqa

It can be seen that, with the identification $\ka_{10}^2\mu_3^2=\ka^2\tau_3^2$, the full contribution to the two point function including 
both the massless NS-NS and R-R exchanges vanishes i.e.,
\beqal{fresult}
\mbox{(\ref{ffinal1})}&+&\mbox{(\ref{fc01})+(\ref{fc21})}=0\non
\mbox{(\ref{ffinal2})}&+&\mbox{(\ref{fc02})+(\ref{fc22})}=0\non
\mbox{(\ref{ffinal3})}&+&\underbrace{\mbox{(\ref{fc03})+(\ref{fc23})}}=0\\
\mbox{NS-NS}&+&\mbox{\hspace{.2in}R-R}\nn
\eeqa

We know from the one loop string calculation that the one loop two point amplitude vanishes (\ref{sum}). In the closed string picture, this cancellation takes place for every mass-level between the NS-NS and R-R states. We will consider similar exchanges for the massless closed strings on the $C^2/Z_2$ orbifold in the next section.

\subsection{Type IIB on $C^2/Z_2$ orbifold}\label{orbifold}

We now turn to the massless closed string exchanges on the orbifold that we are ultimately interested in. 
The procedure followed is same as that of the earlier section. We will first write down the supergravity action on the $C^2/Z_2$ orbifold. We then derive the couplings of the massless closed string modes to the gauge field on a fractional $D_3$. 
In this section we shall primarily make use of the fact that the $Z_2$ orbifold is the singular limit of a smooth ALE space known as Eguchi-Hanson space 
\cite{Eguchi:1978gw}. 
The metric for this space is given by,

\beqal{eh2}
ds^2=f(r)^{-1}dr^2+r^2f(r)\s_z^2+r^2\left[\s_x^2+\s_y^2\right]
\eeqa

\noi where,

\beqa
f(r)=\left[1-\left(\f{a}{r}\right)^4\right] \mbox{\hspace{0.1in}and\hspace{0.1in}} \s_x=-\f{1}{2}\left(\cos{\psi}d\theta+\sin{\theta}\sin{\psi}d\phi\right)\non
\s_y=\f{1}{2}\left(\sin{\psi}d\theta-\sin{\theta}\cos{\psi}d\phi\right)
\mbox{\hspace{0.1in}} \s_z=-\f{1}{2}\left(d\psi+\cos{\theta}d\phi\right) \non
\eeqa

\noi
There is an apparent singularity at $r=a$ which is removed if one identifies
the range of $\psi$ to be $0\le\psi\le 2\pi$. $\theta$ and $\phi$ have the ranges,
$0\le \theta \le \pi$ and $0\le \phi \le 2\pi$.
The space near $r=a$ is locally
$R^2\times S^2$. This is seen from the change variables to $u^2=r^2\left[1-\left(\f{a}{r}\right)^4\right]$, so that for $r=a$ or $u=0$

\beqa
ds^2 \sim \f{1}{4}du^2+\f{1}{4}u^2\left(d\psi +\cos{\theta}d\phi\right)^2
+\f{a^2}{4}\left(d\theta^2+\sin^2{\theta}d\phi^2\right)
\eeqa

\noi
Note that the $R^2$ shrinks to a point as $u \rightarrow 0$. As $r \rightarrow \infty$ the constant $r$ hypersurfaces are given by $S^3/Z_2$. This is due to the fact that the periodicity of $\psi$ here is $2\pi$ instead of the usual periodicity $4\pi$ that gives $S^3$.
The orbifold singulatity arises as the radius of the compact 2-sphere reduces to zero size. The compact 2-sphere (${\cal C}_1$) has an associated antiself-dual two form, $\omega_2$ that is dual to ${\cal C}_1$ and is given by,

\beqa
\omega_2&=&\f{a^2}{2\pi}d\left(\f{\s_z}{r^2}\right)\non
&=& \f{a^2}{2\pi r^3}dr\w d\psi+\f{a^2}{2\pi r^3}\cos{\theta}dr\w d\phi+
\f{a^2}{4\pi r^2}\sin{\theta}d\theta\w d\phi
\eeqa

\noi
This two-form, $\omega_2$ satisfies,

\beqal{conv}
\omega_2=-*\omega_2 \mbox{\hspace{.2in}} \int_{{\cal C}_1}\omega_2=1 \mbox{\hspace{.2in}} \int_{C^2/Z_2}*\omega_2\w \omega_2=\f{1}{2}
\eeqa

\noi
Although the cycle ${\cal C}_1$ shrinks to zero size a non-zero two-form flux $\hat{B}$ persists. A ($p+2$)-form may be decomposed as,
$A_{p+2}=\tilde{A}_p\w \omega_2$. Where $\tilde{A}_p$ is a $p$-form in the transverse six dimensions. This field is twisted and is localised at the orbifold point.
For our analysis we also turn on the ($B+b$) field along the non-orbifolded directions, so that the background is given by

\beqal{matrix}
{\cal B}=
\left(\begin{tabular}{cccc|ccl}
0&1&2&3&...&8&9\\
&&&&&\\
&&$2\pi\al(B+b)$&&&&\\
&&&&&&\\
\hline
&&&&&&\\
&&&&&&$\hat{B}$\\
\end{tabular}\right)
\eeqa

\noi
With the above observations and using equations (\ref{conv}), we can now write down the supergravity action on the orbifold for the twisted fields,

\beqal{actorb}
S_{orb}=-\f{1}{8\ka^2}\int_6 d\tilde{b}\w * d\tilde{b}-\f{1}{8\ka_{10}^2} \int_6 \left[ d\tilde{C_0}\w *d\tilde{C_0}+d\tilde{C_2}\w *d\tilde{C_2}\right]
\eeqa

\noi
Where $\tilde{b}$ is the twisted NS-NS scalar that arises from the dimensional reduction of the $\hat{B}$ so that $\hat{B}=\hat{b}\omega_2$ and,

\beqa
\hat{b}
= 4\pi^2\al\left(\f{1}{2}+\f{\tilde{b}}{4\pi^2\al}\right)
\eeqa

\noi
$\tilde{b}$ is the fluctuating part of $\hat{b}$. Similarly the scalar, $\tilde{C}_0$ and the two-form field $\tilde{C}_2$ arises from the dimensional reduction of the R-R fields $C_2$ and $C_4$ respectively. The propagators for these twisted fields can be easily read off from (\ref{actorb}),

\beqal{ptb}
\expt{\tilde{b}\tilde{b}}=-4i\ka^2\f{1}{\kpe^2+g^{ij}\kpai\kpaj} 
\eeqa

\beqal{ptc0}
\expt{\tilde{C_0}\tilde{C_0}}-4i\ka_{10}^2\f{1}{\kpe^2+g^{ij}\kpai\kpaj} 
\eeqa

\beqal{ptc2}
\expt{\tilde{C_2}\tilde{C_2}}=-4i\ka_{10}^2\f{g_{I[J{'}}g_{I^{'}]J}}
{\kpe^2+g^{ij}\kpai\kpaj}
\eeqa

\subsubsection{NS-NS exchange}\label{ons}

We will now derive the couplings of the gauge field to the twisted NS-NS scalar $\tilde{b}$ that arises from the dimensional reduction of the two form field $\hat{B}$.
In the picture outlined in the begining of Section (\ref{orbifold}), we can view a fractional $D_p$ brane as $D_{p+2}$ brane wrapped on the shrinking cycle ${\cal C}_1$. The Born-Infeld action for a $D_{p+2}$ is 

\beqal{odbi}
S_{p+2}&=&-\tau_{p+2}\int d^{p+3}\xi e^{\f{p-1}{4}\phi}\sqrt{g+{\cal B}e^{-\f{\phi}{2}}}
\eeqa

\noi
where ${\cal B}$ is given by (\ref{matrix}). We can rewrite (\ref{odbi}) as,

\beqal{odbi1}
S_p&=&-\tau_{p+2}\int d^{p+1}\xi e^{\f{p-3}{4}\phi}\sqrt{g+2\pi\al(B+b)e^{-\f{\phi}{2}}}
\int d^2\xi_{\mbox{int}}\sqrt{\hat{B}}\non
&=&-\tau_p\int d^{p+1}\xi e^{\f{p-3}{4}\phi}\sqrt{g+2\pi\al(B+b)e^{-\f{\phi}{2}}}
\left(\f{1}{2}+\f{\tilde{b}}{4\pi^2\al}\right)
\eeqa

\noi
where

\beqal{redb}
\int d^2\xi_{\mbox{int}}\sqrt{\hat{B}}=\int_{{\cal C}_1} \hat{B}
= 4\pi^2\al\left(\f{1}{2}+\f{\tilde{b}}{4\pi^2\al}\right)
\eeqa

\noi
In the second line of (\ref{odbi1}) we have identified,

\beqa
\tau_p=\tau_{p+2}(4\pi^2\al)
\eeqa

\noi
Apart from the usual couplings of the untwisted NS-NS modes, the action gives 
the coupling of the twisted field $\tilde{b}$. 
For the untwisted states the couplings and the two point functions are the 
same as those computed in Section (\ref{fns}) upto an overall constant. 
Here we 
will only be concerned with the twisted field $\tilde{b}$. 
We can write down the coupling of this field to the gauge field $b$ by 
expanding (\ref{odbi}) with various limits of $g$,

\beqa
V_{b\tilde{b}}=\f{2\pi\al}{4\pi}\tau_3 B^{kl} \mbox{\hspace{.2in}(For small $B$ and $g=\eta$)}
\eeqa

\beqa
V_{b\tilde{b}}=\f{\sqrt{2\pi\al B}}{4\pi^2\al}\tau_3\left[\f{1}{2}\left(\f{1}{B}\right)^{kl}+
\f{\ep^2}{2(2\pi\al)^2}\left[\left(\f{1}{B^3}\right)^{kl}-\f{1}{4}\left(\f{1}{B}\right)^{kl}\Tr\left(\f{1}{B^2}\right)\right]\right]\non
\mbox{\hspace{.1in}(For $g=\ep\eta$)}\non
\eeqa

\beqa
V_{b\tilde{b}}=\f{\sqrt{2\pi\al B}}{4\pi^2\al}\tau_3\left[\f{1}{2}\left(\f{1}{B}\right)^{kl}+
\f{(2\pi\al)^2}{2}\left[B^{kl}-\f{1}{4}\left(\f{1}{B}\right)^{kl}\Tr\left({B^2}\right)\right]\right]\non
\mbox{\hspace{.1in}(For $G=\eta$)}\non
\eeqa

\noi
The two point amplitudes with the couplings defined above and the propagator
(\ref{ptb}) are,

\noi
For small $B$,

\beqal{otb1}
L_2(b\tilde{b}b)=-\f{i}{4\pi^2}\ka^2\tau_3^2(2\pi\al)^2\int \f{d^2\kpe}{(2\pi)^2}
\f{1}{\kpe^2+p^2}\left[\f{1}{2}B^{kl}B^{k^{'}l^{'}}+(kl) \leftrightarrow (k^{'}l^{'})\right]
\eeqa

\noi
For $g=\ep\eta$,

\beqal{otb2}
L_2(b\tilde{b}b)&=&-\f{i}{4\pi^2(2\pi\al)^2}\det(2\pi\al B)\ka^2\tau_3^2\int \f{d^2\kpe}{(2\pi)^2}
\f{1}{\kpe^2+\ep^{-1}p^2}\times\non&\times&
\left[{\cal O}(1)+{\cal O}(\ep^2)\right]
\eeqa

\noi
where,

\beqa
{\cal O}(1)=
\left[\f{1}{2}
\left(\f{1}{B}\right)^{kl}\left(\f{1}{B}\right)^{k^{'}l^{'}}
+(kl) \leftrightarrow (k^{'}l^{'})\right]
\eeqa

\beqa
{\cal O}(\ep^2)&=&
\f{\ep^2}{(2\pi\al)^2}\left[\left(\f{1}{B^3}\right)^{kl}
-\f{1}{4}\Tr\left(\f{1}{B^2}\right)\left(\f{1}{B}\right)^{kl}\right]
\left(\f{1}{B}\right)^{k^{'}l^{'}}\non
&+& (kl) \leftrightarrow (k^{'}l^{'})
\eeqa

\noi
For $G=\eta$,

\beqal{otb3}
L_2(b\tilde{b}b)&=&-\f{i}{4\pi^2(2\pi\al)^2}\det(2\pi\al B)\ka^2\tau_3^2\int \f{d^2\kpe}{(2\pi)^2}
\f{1}{\kpe^2+\tilde{p}^2/(2\pi\al)^2}\times\non&\times&\left[{\cal
O}(1)
+{\cal O}(\als)\right]
\eeqa

\beqa
{\cal O}(1)=
\left[\f{1}{2}
\left(\f{1}{B}\right)^{kl}\left(\f{1}{B}\right)^{k^{'}l^{'}}
+(kl) \leftrightarrow (k^{'}l^{'})\right]
\eeqa

\beqa
{\cal O}(\als)&=&
(2\pi\al)^2\left[B^{kl}
-\f{1}{4}\Tr(B^2)\left(\f{1}{B}\right)^{kl}\right]
\left(\f{1}{B}\right)^{k^{'}l^{'}} \non
&+& (kl) \leftrightarrow (k^{'}l^{'})
\eeqa

\subsubsection{R-R exchange}

Similar to the derivation in Section (\ref{ons}), we now expand the 
Chern-Simons action in terms of the twisted and the untwisted R-R fields. 
Keeping in mind the background two-form field (\ref{matrix}) and the 
relations (\ref{conv}). We start with the action for a $D_5$ brane 
wrapping a two cycle ${\cal C}_1$. 

\beqa
S_{CS}&=&i\mu_5\int_{6}\sum_nC_n\w e^{{\cal B}}\non
&=& i\mu_5\f{1}{2}(4\pi^2\al)\left[(2\pi\al)^2\int_4 C_0 B\w b + (2\pi\al)\int_4 C_2\w b\right]\non
&+&i\mu_5\left[(2\pi\al)^2\int_4 \tilde{C}_0 B\w b + (2\pi\al)\int_4 \tilde{C}_2\w b \right]
\eeqa

\noi
Identifying $\mu_3=\mu_5(4\pi^2\al)$,

\beqa
S_{CS}&=&
i\mu_3\f{1}{2}\left[(2\pi\al)^2\int_4 C_0 B\w b + (2\pi\al)\int_4 C_2\w b\right]\non
&+&i\mu_3\f{1}{4\pi^2\al}\left[(2\pi\al)^2\int_4 \tilde{C}_0 B\w b + (2\pi\al)\int_4 \tilde{C}_2\w b \right]
\eeqa
\\
\\
\noi
Note that the twisted and untwisted R-R couplings are same as those computed in Section (\ref{frr})  except for the change in the overall normalisations.
The R-R exchanges for the twisted states thus have the same tensor structures 
as those in Section (\ref{frr}). Incorporating these changes the two point function with twisted R-R exchanges can be written as follows,
\\
\\
\noi
\underline{{\it $\tilde{C_2}$ exchange}} :
\\
\\
For small $B$,

\beqal{oc21}
L_2(b\tilde{C_2}b)&=&\f{i}{{4\pi^2}}\ka_{10}^2\mu_3^2\int
\f{d^2\kpe}{(2\pi)^2}\f{1}{\kpe^2+p^2}
\left[\f{1}{2}\left(\eta^{kk^{'}}\eta^{ll^{'}}-\eta^{kl^{'}}\eta^{lk^{'}}\right)\right]\non
&+& (kl) \leftrightarrow (k^{'}l^{'})
\eeqa

For $g=\ep\eta$,

\beqal{oc22}
L_2(b\tilde{C_2}b)=
&=&\f{i}{{4\pi^2}}\mbox{det}(2\pi\al
B)\ka_{10}^2\mu_3^2\int
\f{d^2\kpe}{(2\pi)^2}\f{1}{\kpe^2+\ep^{-1}p^2}
\times \non
&\times&\f{\ep^2}{(2\pi\al)^4}[\left[\left(\f{1}{B^3}\right)^{lk}
-\f{1}{4}\Tr\left(\f{1}{B^2}\right)\left(\f{1}{B}\right)^{lk}\right]
\left(\f{1}{B}\right)^{l^{'}k^{'}} \non
&+&\f{1}{2}\left[\left(\f{1}{B^2}\right)^{kk^{'}}\left(\f{1}{B^2}\right)^{ll^{'}}
-\left(\f{1}{B^2}\right)^{k^{'}l}\left(\f{1}{B^2}\right)^{kl^{'}}\right]
\non
&+& (kl) \leftrightarrow (k^{'}l^{'})]
\eeqa

For $G=\eta$

\beqal{oc23}
L_2(b\tilde{C_2}b)&=&
\f{i}{{4\pi^2}}\mbox{det}(2\pi\al B)\ka_{10}^2\mu_3^2
\int \f{d^2\kpe}{(2\pi)^2}
\f{1}{\kpe^2+\tilde{p}^2/(2\pi\al)^2}\times\non &\times&
\left[\left[B^{kl}-\f{1}{4}\Tr(B^2)\left(\f{1}{B}\right)^{kl}\right]
\left(\f{1}{B}\right)^{k^{'}l^{'}}
+\f{1}{2}\left(\eta^{ll^{'}}\eta^{kk^{'}}
-\eta^{kl^{'}}\eta^{lk^{'}}\right)\right]\non &+& (kl) \leftrightarrow
(k^{'}l^{'})
\eeqa
\\
\newpage
\noi
\underline{{\it $\tilde{C_0}$ Exchange}}:
\\
\beqal{oc01}
L_2(b\tilde{C_0}b)&=&\f{i}{{4\pi^2}}\ka_{10}^2\mu_3^2 (2\pi\al)^2 \int\f{d^2\kpe}{(2\pi)^2}
\f{1}{\kpe^2+p^2}\times\\&\times&
[\f{1}{2}B^{kl}B^{k^{'}l^{'}}-\f{1}{4}\Tr(B^2)\left(\eta^{kk^{'}}\eta^{ll^{'}}-\eta^{kl^{'}}\eta^{lk^{'}}\right)
\non &+&\left[\eta^{kk^{'}}(B^2)^{ll^{'}}-\eta^{kl^{'}}(B^2)^{lk^{'}}
\right]+(kl) \leftrightarrow (k^{'}l^{'})]
\mbox{\hspace{.1in} (for small $B$)}\nn
\eeqa

\beqal{oc02}
L_2(b\tilde{C_0}b)&=&\f{i}{{4\pi^2}(2\pi\al)^2}\mbox{det}(2\pi\al)\ka_{10}^2\mu_3^2\int \f{d^2\kpe}{(2\pi)^2}
\f{1}{\kpe^2+\ep^{-1}p^2}\times\non&\times&\left[\f{1}{2}
\left(\f{1}{B}\right)^{kl}\left(\f{1}{B}\right)^{k^{'}l^{'}}
+(kl) \leftrightarrow (k^{'}l^{'})\right]\mbox{\hspace{.1in} (for $g=\ep\eta$)}
\eeqa

\beqal{oc03}
L_2(b\tilde{C_0}b)&=&\f{i}{{4\pi^2}(2\pi\al)^2}\mbox{det}(2\pi\al)\ka_{10}^2\mu_3^2\int \f{d^2\kpe}{(2\pi)^2}
\f{1}{\kpe^2+\tilde{p}^2/(2\pi\al)^2}\times\non&\times&\left[\f{1}{2}
\left(\f{1}{B}\right)^{kl}\left(\f{1}{B}\right)^{k^{'}l^{'}}
+(kl) \leftrightarrow (k^{'}l^{'})\right] \mbox{\hspace{.1in} (for $G=\eta$)}
\eeqa

We have seen that the untwisted exchanges for both the NS-NS and R-R sectors 
are the same as those computed in section (\ref{flat}) modulo an overall 
normalisation. The sum of these thus vanishes just like the flat case, 
eqns(\ref{fresult}). This is also what we get from the one loop computation. 
See eqn(\ref{sum}). The twisted states however sum up to finite results. We 
write these contributions below with the identification 
$\ka_{10}\mu_3=\ka\tau_3$,

\beqa
L_2&=&\mbox{(\ref{otb1})+(\ref{oc21})+(\ref{oc01})}\mbox{\hspace{.1in} (for small $B$)}\non
&=&\f{i}{{4\pi^2}}\ka_{10}^2\mu_3^2  \int\f{d^2\kpe}{(2\pi)^2}
\f{1}{\kpe^2+p^2}\times\non&\times&
[\f{1}{2}\left[1-(2\pi\al)^2\f{1}{2}\Tr(B^2)\right]\left(\eta^{kk^{'}}\eta^{ll^{'}}-\eta^{kl^{'}}\eta^{lk^{'}}\right)
\non &+&(2\pi\al)^2\left[\eta^{kk^{'}}(B^2)^{ll^{'}}-\eta^{kl^{'}}(B^2)^{lk^{'}}
\right]+(kl) \leftrightarrow (k^{'}l^{'})] 
\eeqa

\beqa
L_2&=&\mbox{(\ref{otb2})+(\ref{oc22})+(\ref{oc02})}\mbox{\hspace{.1in} (for $g=\ep\eta$)}\non
&=&\f{i}{{4\pi^2}}\mbox{det}(2\pi\al
B)\ka_{10}^2\mu_3^2\int
\f{d^2\kpe}{(2\pi)^2}\f{1}{\kpe^2+\ep^{-1}p^2}
\times \non
&\times&\f{\ep^2}{(2\pi\al)^4}
\f{1}{2}\left[\left(\f{1}{B^2}\right)^{kk^{'}}\left(\f{1}{B^2}\right)^{ll^{'}}
-\left(\f{1}{B^2}\right)^{k^{'}l}\left(\f{1}{B^2}\right)^{kl^{'}}\right]
\non &+& (kl) \leftrightarrow (k^{'}l^{'})
\eeqa

\beqa
L_2&=&\mbox{(\ref{otb3})+(\ref{oc23})+(\ref{oc03})}\mbox{\hspace{.1in} (for $G=\eta$)}\non
&=&
\f{i}{{4\pi^2}}\mbox{det}(2\pi\al B)\ka_{10}^2\mu_3^2
\int \f{d^2\kpe}{(2\pi)^2}
\f{1}{\kpe^2+\tilde{p}^2/(2\pi\al)^2}\times\non &\times&
\f{1}{2}\left[\eta^{ll^{'}}\eta^{kk^{'}}
-\eta^{kl^{'}}\eta^{lk^{'}}\right]\non &+& (kl) \leftrightarrow
(k^{'}l^{'})
\eeqa

These are the exact expansions of the one loop sting amplitude, 
eqn(\ref{2ptfinal}) when $G$ and $\det(g+2\pi\al B)$ are expanded to 
respective orders as follows,

\beqa
g&=&\eta\non
G^{ij}&\sim & \eta^{ij}+(2\pi\al)^2(B^2)^{ij}+{\cal O}(B^4)\non
\sqrt{\eta+(2\pi\al)B}&\sim &\left[1-\f{(2\pi\al)^2}{4}\Tr(B^2)
+{\cal O}(B^4)\right]
\eeqa

\beqa
g&=&\ep\eta \non
G^{ij}&\sim &-\f{\ep}{(2\pi\al)^2}\left(\f{1}{B^2}\right)^{ij} +{\cal
O}(\ep^3)\non
\sqrt{\ep\eta+(2\pi\al)B}&\sim &
\sqrt{(2\pi\al)B}\left[1-\f{\ep^2}{4(2\pi\al)^2}\Tr\left(\f{1}{B^2}\right)
\right]
\eeqa

\beqa
G&=&\eta\non
g&=&-(2\pi\al)^2B^2 +{\cal O}(B^4)\non
\sqrt{g+(2\pi\al)B}&\sim&
\sqrt{(2\pi\al)B}\left[1-\f{(2\pi\al)^2}{4}\Tr(B^2)\right]
\eeqa

\noi
The computation of the two point amplitude in string theory sums up all the $B$-field dependence in the open string metric $G$ and $\det(g+2\pi\al B)$. The analysis in this section reproduces these terms to the orders relevant in the expansion about the various limits of the closed string metric $g$.

\section{Conclusion}

In this paper we have studied closed string exchanges in the presence of a 
background $B$-field. The main aim is to identify the propagating massless 
closed string modes that contribute to the non-zero two point amplitude. We 
have already seen from the one loop string computation in \cite{myopcl2}, 
that for the noncommutative ${\cal N}=2$ gauge theory on a fractional $D_3$ 
brane, the tree-level massless closed string exchanges give the same result 
as the one loop gauge theory.
Thus the IR divergent terms that arise by integrating over high energy modes 
in the loops in the nonplanar amplitudes can be interpreted as coming from the 
massless closed string modes. This follows as a consequence of the world-sheet 
open closed string duality. This correspondence between the lowest lying modes 
on either channel is possible as a consequence of cancellation of all massive 
modes contributing to the loop diagram. 
Here we have reproduced this amplitude by considering massless closed string 
exchanges from the effective field theory. Specifically for a fractional $D_3$ brane localised at the fixed point of $C^2/Z_2$ orbifold, the closed strings that give finite amplitude come from the twisted NS-NS and R-R sectors. The contributions from the untwisted NS-NS and R-R modes cancel amongst themselves. The duality discussed in the paper was already known for the commutative gauge theory \cite{douglas2}. We have seen that the background $B$-field primarily acts as a regulator for the non-planar diagrams and preserves this duality. On can thus expect to recover the UV/IR divergences of the noncommutative gauge theory from a finite number massless closed string exchanges in models where the commutative theory has this duality at least perturbatively. These cases have recently been studied. (See \cite{DiVecchia:2005vm} for a review and references). \\

\noindent
{\bf Acknowledgements :} 
I am indebted to B. Sathiapalan for various comments, suggestions and 
for carefully reading the manuscript. I would also like to thank T. K. Dey, 
A. Kumar, S. Mukherji and B. Rai for discussions.

\end{document}